\documentstyle[seceq,preprint,wrapfig]{ptptex}

\def\KK{$K^0$-$\bar{K^0}$}
\def\BB{$B^0$-$\bar{B^0}$}
\def\BdBd{$B_d^0$-$\bar{B_d^0}$}
\def\BsBs{$B_s^0$-$\bar{B_s^0}$}

\def\Bd{B_d^0}
\def\Bs{B_s^0}
\def\e{\epsilon}
\def\DM{\Delta M}
\def\w{\tilde\omega}
\def\x{\tilde\chi}
\def\g{\tilde g}
\def\sq{\tilde q}
\def\lsim{\ ^<\llap{$_\sim$}\ }
\def\gsim{\ ^>\llap{$_\sim$}\ }
\def\r2{\sqrt 2}

\def\sw2{\sin^2\theta_W}
\def\v#1{v_#1}
\def\tb{\tan\beta}

\def\s2b{\sin 2\beta}
\def\c2b{\cos 2\beta}
\def\s2b2{\sin^22\beta}
\def\uL{{\tilde u}_L}
\def\uR{{\tilde u}_R}
\def\cL{{\tilde c}_L}
\def\cR{{\tilde c}_R}
\def\tL{{\tilde t}_L}
\def\tR{{\tilde t}_R}

\def\MU2{M_U^2}
\def\MCH{M_{H^\pm}}
\def\mgr{m_{3/2}}
\def\m#1{{\tilde m}_#1}
\def\mH{m_H}

\def\mw#1{{\tilde m}_{\omega #1}}
\def\mQ{{\tilde m}_Q}
\def\mU{{\tilde m}_U}
\def\MuSL{{\tilde M}_{uL}^2}
\def\MuSR{{\tilde M}_{uR}^2}
\def\McSL{{\tilde M}_{cL}^2}
\def\McSR{{\tilde M}_{cR}^2}

\def\MtS#1{{\tilde M}_{t#1}^2}

\preprintnumber{
 OCHA-PP-68}

\title{%
Signatures of Supersymmetry at B-Factories
\footnote{to be published in the proceedings of the
Yukawa International Seminar '95.}
}

\author{%
Noriyuki {\sc Oshimo}\footnote{
E-mail address: oshimo@phys.ocha.ac.jp}
}

\inst{%
Department of Physics, Ochanomizu University, Tokyo 112
}

\recdate{%
\today
}

\abst{%
     We discuss $CP$ asymmetries in
$B^0$-meson decays and \BsBs\ mixing
within the framework of the supersymmetric
standard model (SSM).  It is shown that
\BdBd\ and \KK\ mixings could receive sizable new
contributions through box diagrams mediated by the
charginos and charged Higgs bosons.
This implies that the $CP$-violating phase
of the Cabibbo-Kobayashi-Maskawa matrix may have a value which
is different from the one predicted by the
standard model (SM).
The value of the $CP$-violating phase affects the amounts of
the $CP$ asymmetries and \BsBs\ mixing.
We examine predictions for these quantities
in both the SSM and the SM, exploring the difference
between the two models.
}

\begin{document}

\maketitle

\section{Introduction}

     In the supersymmetric standard model
(SSM) there exist several new sources
for flavor-changing neutral current
(FCNC) processes, such as \BB\ mixing and radiative $B$-meson
decay. The new sources are
the interactions \cite{ellis} in which a quark $q$ couples to a
squark $\sq$ of a different generation
and a chargino $\w$, a neutralino $\x$,
or a gluino $\g$.  Since the SSM
contains two doublets of Higgs bosons, charged
Higgs bosons $H^{\pm}$ also mediate FCNC processes \cite{wise}.

     In this article we discuss SSM contributions
to \BdBd\ and \KK\ mixings \cite{BB} and their effects on
$CP$ asymmetries in neutral $B$-meson decays and \BsBs\ mixing
\cite{bb2},
searching for signatures of supersymmetry.
The measurements of these quantities
will be performed at B-factories in the near future.
Possible observation of discrepancies within
standard model (SM) predictions can give hints for physics
beyond the SM.  Since the SSM is one of the most plausible
extensions of the SM, studying the predictions of the SSM
would be important.
It will be shown that SSM predictions are deviated from
SM predictions in sizable ranges of SSM parameters.
For definiteness, we assume
the framework of grand unification
theories coupled to $N=1$ supergravity \cite{SUSY}.

     Among the new sources for FCNC processes in the SSM, sizable
contributions can be
expected from the interactions mediated by the
charginos and the charged Higgs bosons \cite{BB,bsg,SUSYBB}.
The reasons are as follows:  Since the $t$-quark has a
large mass, the coupling strengths of the related
Yukawa interactions for the chargino, the $t$-squark, and
the down-type quark and for the charged Higgs boson, the $t$-quark,
and the down-type quark are comparable to that of
the SU(2) gauge interaction.  Consequently,
the chargino interaction strengths for the $t$-squarks
are made different from those for the $u$- or $c$-squarks
which are determined by the gauge interaction alone.
Besides, if the squark masses of the first two generations are
not much different from the $t$-quark mass, one
of the $t$-squarks can be lighter than the other
squarks.  These effects soften the cancellation among
different squark contributions for the chargino
box diagrams, which otherwise is rather severe.
In the box diagrams exchanging charged Higgs bosons,
the charged Higgs boson interactions for the $t$-quark
are no longer weak, compared to the standard $W$-boson
interactions, while those for the $u$- or $c$-quark
are much weaker.  Therefore, the chargino and the
charged Higgs boson contributions could
naturally be the same order of magnitude as the
$W$-boson contributions.  On the other hand, the gluino
and the neutralino contributions are mediated by
the down-type squarks, whose masses are quite
degenerated.  Although the interaction strength for
the gluinos is stronger than for the charginos, in the
grand unification scheme the gluino mass is proportionally
larger than the chargino masses.
Therefore, the gluino and the neutralino contributions
become smaller than the chargino contributions.

     In sect. 2 we briefly review the model.  In sect. 3
we discuss the SSM contributions to \BdBd\ and \KK\
mixings.  In sect. 4 we evaluate
the Cabibbo-Kobayashi-Maskawa
(CKM) matrix through these mixings and discuss the resultant
implications for $CP$ asymmetries in $B^0$-meson decays and
\BsBs\ mixing.
Summary
is given in sect. 5.

\section{Model}

     In the SSM, down-type quarks interact with charginos and
up-type squarks.
The charginos are the mass eigenstates of
the SU(2) charged gauginos and the charged higgsinos.
Their mass matrix is given by
\begin{equation}
    M^- = \left(\matrix{\m2 & -{1\over\r2}g\v1 \cr
                -{1\over\r2}g\v2 & \mH}        \right),
\label{ (1)}
\end{equation}
where $\v1$ and $\v2$ are the vacuum expectation values
of the Higgs bosons, and $\m2$ and $\mH$ respectively
denote the SU(2) gaugino mass and the higgsino mass
parameter.  In the ordinary scheme for generating the gaugino
masses, $\m2$ is smaller than or around the
gravitino mass $\mgr$.  If the SU(2)$\times$U(1) symmetry
is broken through radiative corrections,
a relation $\tb$ $(\equiv \v2/\v1 ) \gsim 1$
holds and the magnitude of $\mH$ is at most of order of $\mgr$.
The chargino mass eigenstates are obtained by
diagonalizing the matrix $M^-$ as
\begin{equation}
      C_R^\dagger M^-C_L = {\rm diag}(\mw1, \mw2) \quad
                       (\mw1 <\mw2 ),
\label{ (2)}
\end{equation}
$C_R$ and $C_L$ being unitary matrices.

     The squark fields, as well as the quark fields, are
mixed in generation
space.  Since the left-handed
squarks and the right-handed ones are also mixed,
the mass-squared matrix for the
up-type squarks $\MU2$ is expressed by a $6\times 6$ matrix:
\begin{eqnarray}
    \MU2 &=& \left(\matrix{{\MU2}_{11} & {\MU2}_{12} \cr
                        {\MU2}_{21} & {\MU2}_{22} }    \right);
                                    \nonumber \\
{\MU2}_{11}&=&\c2b ({1\over 2}-{2\over 3}\sw2 )M_Z^2 + {\mQ}^2
            + (1+c)m_U m_U^{\dagger},
                           \nonumber \\
{\MU2}_{22}&=&{2\over 3}\c2b\sw2 M_Z^2 + {\mU}^2
            + (1+2c)m_U^{\dagger}m_U,  \nonumber \\
{\MU2}_{12}&=&{\MU2}_{21}^{\dagger} =\cot\beta\mH m_U
                                             + a^*\mgr m_U,
\label{eq.3}
\end{eqnarray}
where $m_U$ denotes the mass matrix of the
up-type quarks.  The mass parameters
$\mgr$, $\mQ$, $\mU$ and the dimensionless constants
$a$, $c$ come from the terms in the SSM Lagrangian
which break supersymmetry softly:
$\mQ$ and $\mU$ are determined by the gravitino
and gaugino masses and $\mQ\simeq\mU\sim\mgr$;
$a$ is related to the breaking of local supersymmetry;
and $c$ represents the magnitude of radiative corrections to the
squark masses.  At the electroweak energy scale, $a$
is of order unity and $c=-1-(-0.1)$.

     The squark mixings
among different generations in Eq. (\ref{eq.3}) are
removed by a unitary matrix which consists of
the same $3\times 3$ matrices that diagonalize the
up-type quark mass matrix \cite{bsg}.
As a result,
the generation mixings in the Lagrangian between
the down-type quarks and the up-type squarks in
mass eigenstates are described by the CKM
matrix of the quarks.
The mixings between the left-handed and right-handed squarks
can be neglected for the first two generations
because of the smallness of the corresponding quark masses.  The
masses of the left-handed squarks $\uL, \cL$ and the
right-handed squarks $\uR, \cR$ are given by
\begin{eqnarray}
\MuSL =\McSL &=& {\mQ}^2+\c2b({1\over 2}-{2\over 3}\sw2)M_Z^2,
                                   \nonumber \\
\MuSR =\McSR &=& {\mU}^2+{2\over 3}\c2b\sw2 M_Z^2.
\label{(4)}
\end{eqnarray}
For the third generation, the large
$t$-quark mass leads to an appreciable
mixing between $\tL$ and $\tR$.
The mass-squared matrix for the $t$-squarks
becomes
\begin{equation}
M_t^2 = \left(\matrix{\MuSL +(1-|c|)m_t^2 &
                     (\cot\beta\mH +a^*\mgr)m_t \cr
              (\cot\beta\mH^* +a\mgr)m_t &
                \MuSR +(1-2|c|)m_t^2 }    \right).
\label{ (5)}
\end{equation}
The mass eigenstates of the $t$-squarks are obtained by
diagonalizing the matrix $M_t^2$ as
\begin{equation}
  S_tM_t^2S_t^\dagger ={\rm diag}(\MtS1, \MtS2) \quad
                      (\MtS1 <\MtS2),
\label{ (6)}
\end{equation}
where $S_t$ is a unitary matrix.

     Down-type quarks interact also with charged Higgs bosons
and up-type quarks.  The generation mixings in these
interactions are described by the CKM matrix.
The parameters which determine FCNC processes, other than the
SM parameters, are only the charged Higgs boson mass $\MCH$
and $\tb$.

     In general, the SSM parameters have complex
values.  If the magnitudes of their physical complex
phases are of order unity, the electric dipole
moment of the neutron is predicted to have a large value.
Its experimental limits then lead to a lower
bound of about 1 TeV on the squark masses \cite{EDM}.
In this case, the SSM does not give any sizable
new contributions to FCNC processes.
We assume hereafter real values for the
parameters other than the SM parameters, so that
the constraints from the electric dipole moment
of the neutron can be ignored.

\section{\BB\ and \KK\ mixings}

     The SSM gives new contributions
to \BB\ and \KK\ mixings through
box diagrams mediated by the charginos or
the charged Higgs bosons.
For the chargino contribution, exchanged bosons are
up-type squarks.
For the charged Higgs boson contribution,
exchanged fermions are up-type quarks and exchanged
bosons are either only charged Higgs
bosons or charged Higgs bosons and $W$-bosons.

     An observable quantity for
\BdBd\ mixing is the mixing parameter
$x_d =\DM_{B_d}/\Gamma_{B_d}$ \cite{FCNC}, where $\DM_{B_d}$ and
$\Gamma_{B_d}$ denote the mass difference and the average width
for the $B_d^0$-meson mass eigenstates.  The mass difference
is induced dominantly by the short distance contributions of
box diagrams.  The mixing parameter is given by
\begin{equation}
  x_d={G_F^2\over 6\pi^2}M_W^2{M_{B_d}\over \Gamma_{B_d}}
    f_{B_d}^2B_{B_d}|V_{31}^*V_{33}|^2
              \eta_{B_d}|A^W_{tt}+A^C+A^H_{tt}|,
\label{xd}
\end{equation}
where $G_F$, $f_{B_d}$, $B_{B_d}$, and $\eta_{B_d}$
represent the Fermi constant,
the $B_d^0$-meson decay constant, the bag factor for
$B_d^0$-$\bar{B_d^0}$ mixing, and the QCD correction factor.
The CKM matrix is denoted by $V$.  The contributions
of the $W$-boson, chargino,
and charged Higgs boson box diagrams are respectively
expressed as $A^W_{tt}$, $A^C$, and $A^H_{tt}$, which
are explicitly given in Refs. \citen{bb2,inami}.
The box diagrams with $t$-quarks give $A^W_{tt}$ and
$A^H_{tt}$.

     For \KK\ mixing,
the $CP$ violation parameter $\e$
is an observable quantity for the short distance contributions
of box diagrams.
This parameter can be written as
\begin{eqnarray}
  \epsilon &=& -{\rm e}^{i\pi/4}
     {G_F^2\over 12\r2\pi^2}M_W^2{M_K\over \Delta M_K}f_K^2B_K
      {\rm Im}[(V_{31}^*V_{32})^2\eta_{K33}(A_{tt}^W+A^C+A_{tt}^H)
                                        \nonumber \\
       & &+ (V_{21}^*V_{22})^2\eta_{K22}A_{cc}^W
 + 2V_{31}^*V_{32}V_{21}^*V_{22}\eta_{K32}A_{tc}^W],
\label{ep}
\end{eqnarray}
where $f_K$ and $B_K$ represent the decay constant and the
bag factor.  The QCD correction factors are denoted
by $\eta_{Kab}$.
The standard $W$-boson box diagram with $c$-quarks
and that with $t$- and $c$-quarks give
$A^W_{cc}$ and $A^W_{tc}$, respectively.

     The SSM prediction for $x_d$ is different from the
SM prediction by $A^C+A^H_{tt}$.
The difference between
the SSM and the SM predictions for $\e$ is in
the term proportional to $(V_{31}^*V_{32})^2$,
which is the same amount as for $x_d$.  Thus,
we can measure the amount of the SSM contributions to \BdBd\ and
\KK\ mixings by the ratio
\begin{equation}
R={A_{tt}^W+A^C+A_{tt}^H\over A_{tt}^W}.
\label{ratio}
\end{equation}
If new contributions are negligible, $R$ becomes
unity.

\begin{table}
\caption{The ratio $R_C$ for $\m2=200$ GeV and
         $\mQ=200$ GeV.  The other parameters are
         set for $\mU=a\mgr=\mQ$ and $|c|=0.3$:
	 (i) $\tb=1.2$, (ii) $\tb=2$.}
\begin{center}
\begin{tabular}{c|c|c|c|c}
\hline
$\mH$ (GeV)       & $-200$   & $-100$    & 100  & 200       \\
\hline
(i)   & 1.26   & 1.43  &  2.25  &   \\
\hline
(ii)     & 1.14  & 1.25 & 1.50   &  1.35 \\
\hline
\end{tabular}
\end{center}
\label{tableR}
\end{table}
     We now examine the amounts of SSM contributions to
the mixings.
In order to see the chargino and the charged Higgs boson
contributions separately, we evaluate, instead of $R$,
the ratios $R_C=(A^W_{tt}+A^C)/A^W_{tt}$ and
$R_H=(A^W_{tt}+A^H_{tt})/A^W_{tt}$.
In Table \ref{tableR} the value of $R_C$ is given for
$\tb=1.2$ (i), 2 (ii) and several values of the higgsino
mass parameter $\mH$.
The other parameters are set, as typical values,
for $\m2=200$ GeV, $\mQ=200$ GeV, $\mQ=\mU=a\mgr$,
and $|c|=0.3$.  For the $t$-quark mass we use
$m_t=170$ GeV \cite{top}.
In case (i) with $\mH=200$ GeV, the lighter $t$-squark
mass-squared becomes negative.
The sign of
the chargino contribution is the same as that of the
$W$-boson contribution, and these contributions interfere
constructively.
If $\tb$ is not much larger than unity,
the ratio $R_C$ can have a value larger than 1.5
in sizable ranges of other SSM parameters.
The dependence of $R_C$ on $\tb$
arises from the chargino Yukawa interactions:  $R_C$
increases as $\tb$ decreases, since a smaller value for
$v_2$ enhances the Yukawa couplings of the charginos to the
$t$-squarks.
The ratio $R_C$ also depends on $\mQ$, whereas it does
not vary so much with $\m2$.
In Table \ref{tableH} we show $R_H$ for
$\tb=1.2$ (i), 2 (ii) and $\MCH=100$, 200, 300 GeV.
The charged Higgs boson box diagrams also
contribute constructively.  In case (i) $R_H$ is
larger than 1.5 for $\MCH\lsim$ 180 GeV.
Similarly to the chargino contribution,
the value of $R_H$ increases as $\tb$ decreases.
The net amount of the SSM contribution
is given by the sum of all the
contributions.  The ratio $R$ in Eq. (\ref{ratio}) is larger than
$R_C$ or $R_H$.
For example, in case (i) with $\MCH=200$ GeV, the ratio $R$
becomes $R\simeq 1.9$ for $\mH=-100$ GeV
(${\tilde M}_{t1}\simeq 188$ GeV, $\mw1\simeq 119$ GeV)
and $R\simeq 2.7$ for $\mH=100$ GeV
(${\tilde M}_{t1}\simeq 85$ GeV, $\mw1\simeq 56$ GeV),
where
${\tilde M}_{t1}$ and $\mw1$ denote the lighter $t$-squark
mass and the lighter chargino mass, respectively.
\begin{table}
\caption{The ratio $R_H$: (i) $\tb=1.2$, (ii) $\tb=2$.}
\begin{center}
\begin{tabular}{c|c|c|c}
\hline
$\MCH$ (GeV)       &  100  & 200 &300       \\
\hline
(i)   & 1.75   & 1.45   & 1.30  \\
\hline
(ii)     & 1.24  & 1.15  & 1.10   \\
\hline
\end{tabular}
\end{center}
\label{tableH}
\end{table}

\section{$CP$ asymmetries}

     We discuss what an enhancement of $R$ implies for
observable quantities.
The values of $x_d$ and $\e$ have been experimentally measured as
$x_d=0.71\pm 0.06$ and $|\e|=2.26\times 10^{-3}$ \cite{PDG}.
An enhanced value of $R$ is considered to
give a prediction, for the CKM matrix, different
from the SM prediction through Eqs. (\ref{xd}), (\ref{ep}).
This can be seen easily, if we adopt the standard
parametrization \cite{PDG} for the CKM matrix:
\begin{equation}
V = \left(\matrix{c_{12}c_{13} & s_{12}c_{13} &
                    s_{13}e^{-i\delta}  \cr
      -s_{12}c_{23}-c_{12}s_{23}s_{13}e^{i\delta} &
      c_{12}c_{23}-s_{12}s_{23}s_{13}e^{i\delta} &
          s_{23}c_{13}             \cr
      s_{12}s_{23}-c_{12}c_{23}s_{13}e^{i\delta} &
      -c_{12}s_{23}-s_{12}c_{23}s_{13}e^{i\delta} &
          c_{23}c_{13}       }    \right),
\label{ (13)}
\end{equation}
where $c_{ab}=\cos\theta_{ab}$ and $s_{ab}=\sin\theta_{ab}$.
Without loss of generality, the angles $\theta_{12}$,
$\theta_{23}$, and $\theta_{13}$ can be taken to lie
in the first quadrant, leading to $\sin\theta_{ab}>0$ and
$\cos\theta_{ab}>0$.  At present, the experiments give
$|V_{12}|=0.22$, $|V_{23}|=0.04\pm 0.004$, and
$|V_{13}/V_{23}|=0.08\pm 0.02$ \cite{PDG,CKM}
within the framework of the SM.
Since these values have been measured through the
processes for which new contributions by the
SSM, if any, are negligible, they are also valid
in the SSM.  Among the four independent parameters of the
CKM matrix, the value of $\sin\theta_{13}$ is determined by
$\sin\theta_{13}=|V_{13}|$.  Owing to this
smallness of $\sin\theta_{13}$, the values of
$\sin\theta_{12}$ and $\sin\theta_{23}$ are given by
$\sin\theta_{12}=|V_{12}|$ and
$\sin\theta_{23}=|V_{23}|$.  The remaining undetermined
parameter is the $CP$-violating phase $\delta$, which
can be measured by $x_d$ or $\e$.  Therefore,
the value of $\delta$ depends on $R$ and the SSM with $R>1$
and the SM predict different values for it.

     The value of $\delta$ is determined as a function
of $R$ or vice versa independently by $x_d$ and $\e$.
Using consistency in these two evaluations,
we can specify the values of $\delta$ and $R$.
In Table \ref{cosd} we show the allowed range
of $\cos\delta$ derived from
the experimental values of $x_d$, $\epsilon$, and
CKM matrix elements, for several values of the ratio $R$.
The sign of $\sin\delta$ should be positive in order to
give $\e$ correctly.
We have assumed the experimental central value for
$x_d$ and taken three sets of values for the CKM matrix
elements:
$(|V_{13}/V_{23}|, |V_{23}|)=(0.08, 0.04)$ (a),
(0.08, 0.044) (b), (0.1, 0.04) (c).
The theoretical uncertainties
of $B_K$, $B_{B_d}$, and $f_{B_d}$ are incorporated
as 0.6 $<B_K<$ 0.9 from a combined result of
lattice \cite{Klattice} and 1/N \cite{KN} calculations and
180 MeV $<f_{B_d}\sqrt{B_{B_d}}<$ 260 MeV from a lattice
calculation \cite{Blattice}.
For the QCD correction
factors we have used $\eta_{B_d}=0.55$ in Eq. (\ref{xd}) and
$\eta_{K33}=0.57$, $\eta_{K22}=1.1$, and $\eta_{K32}=0.36$
in Eq. (\ref{ep}) \cite{QCD}.
For $R=2.5$ in case (a) and $R=2$, 2.5 in case (b),
there is no allowed range of $\cos\delta$.
If we assume case (a), which corresponds to
experimental central values for
$|V_{13}/V_{23}|$ and $|V_{23}|$,
the allowed ranges are $0.8\lsim R\lsim 2.1$ and
$-0.5\lsim\cos\delta\lsim 0.8$, while
$-0.1\lsim\cos\delta\lsim 0.3$ for the SM of $R=1$.
The ratio $R$ cannot be much larger than 2, which rules
out some regions in the SSM parameter space.
In particular, the existence of a very light $t$-squark
becomes unlikely \cite{stop}.
Within the possible ranges
for $|V_{13}/V_{23}|$ and $|V_{23}|$ taking into account
the experimental uncertainties, the ratio $R$
can be at most $R\sim 3$.

\begin{table}
\caption{The allowed range of $\cos\delta$:
          (a) $|V_{13}/V_{23}|=0.08$, $|V_{23}|=0.04$,
          (b) $|V_{13}/V_{23}|=0.08$, $|V_{23}|=0.044$,
          (c) $|V_{13}/V_{23}|=0.1$, $|V_{23}|=0.04$.}
\begin{center}
\begin{tabular}{c|c|c|c|c}
\hline
$R$       &  1  & 1.5  &  2  &  2.5     \\
\hline
(a)   & ($-0.09$, 0.29)   & (0.46, 0.64) & (0.74, 0.77) &  \\
\hline
(b)   & (0.20, 0.65)   & (0.66, 0.82)  &              &  \\
\hline
(c) & (0.01, 0.51) & (0.45, 0.73) & (0.68, 0.82) & (0.81, 0.88) \\
\hline
\end{tabular}
\end{center}
\label{cosd}
\end{table}
     The value of
$\cos\delta$ in the SSM could be larger than
that allowed in the SM, as seen from Table \ref{cosd}.
For instance, in case (a)
the range
$0.3\lsim\cos\delta\lsim 0.8$ is allowed only in the SSM.
In near future experiments,
$CP$ asymmetries in $B^0$-meson decays
and amount of $B^0_s$-$\bar{B^0_s}$ mixing
will be measured, which depend on $\cos\delta$.  It is
possible that these physical quantities
have values outside the ranges predicted by the SM.

     The $CP$ asymmetries enable to measure the angles of
the unitarity triangle given by
\begin{equation}
  \phi_1 =
   {\rm arg}\left( -{V_{21}V_{23}^*\over V_{31}V_{33}^*}\right) ,
                \quad
  \phi_2 =
   {\rm arg}\left( -{V_{31}V_{33}^*\over V_{11}V_{13}^*}\right) ,
                 \quad
  \phi_3 =
   {\rm arg}\left( -{V_{11}V_{13}^*\over V_{21}V_{23}^*}\right) .
\label{ (15)}
\end{equation}
For instance, the decays
$\Bd\rightarrow\psi K_S$,
$\Bd\rightarrow\pi^+\pi^-$,
and $\Bs\rightarrow\rho K_S$
can be used to determine
$\sin 2\phi_1$, $\sin 2\phi_2$,
and $\sin 2\phi_3$, respectively \cite{CPasy}.
These asymmetries are expressed
in terms of $\cos\delta$ and
$r\equiv \cot\theta_{12}(\sin\theta_{13}/\sin\theta_{23})$,
to an excellent accuracy:
\begin{eqnarray}
  \sin 2\phi_1 &=&
\frac{2r\sin\delta(1-r\cos\delta)}{1-2r\cos\delta +r^2}, \quad
  \sin 2\phi_2 =
\frac{2\sin\delta(r-\cos\delta)}{1-2r\cos\delta +r^2},
                             \nonumber \\
  \sin 2\phi_3 &=& \sin 2\delta.
\label{(21)}
\end{eqnarray}
The values of $\sin 2\phi_1$ and $\sin 2\phi_2$
only depend on $\cos\delta$ and $r$, while $\phi_3=\delta$.
In most of the plausible ranges for $\cos\delta$
inferred from the analyses of $x_d$ and $\e$,
the values of $\sin 2\phi_2$ and
$\sin 2\phi_3$ monotonously change between $-1$ and 1,
while $\sin 2\phi_1$ does not vary much with $\cos\delta$.
The dependence on $|V_{13}/V_{23}|$ is weak for both
$\sin 2\phi_1$ and $\sin 2\phi_2$.
These show the reasons why, in the SM, the prediction of
$\sin 2\phi_1$ has been made more specifically
than those of
$\sin 2\phi_2$ and $\sin 2\phi_3$ \cite{gilman}.
Taking into account the present constraints on $\cos\delta$
and $|V_{13}/V_{23}|$, the value of
$\sin 2\phi_1$ should satisfy
$0.4\lsim\sin 2\phi_1\lsim 0.8$, whereas
$\sin 2\phi_2$ and $\sin 2\phi_3$ can have
any values from $-1$ to 1.

     The mixing parameter $x_s$ for $B^0_s$-$\bar{B^0_s}$ mixing
is given by an equation analogous to Eq. (\ref{xd}).
The ratio of $x_s$ to $x_d$ becomes
\begin{equation}
     {x_s\over x_d}={|V_{32}|^2\over |V_{31}|^2},
\label{eq.16}
\end{equation}
where we have neglected the small differences between
$\Bd$ and $\Bs$ caused by the SU(3)$_{flavor}$ breaking.
This ratio is expressed in terms of $\cos\delta$ and $r$ as
\begin{equation}
    \frac{x_s}{x_d}=\frac{\cot^2\theta_{12}+2r\cos\delta}
                         {1-2r\cos\delta+r^2}.
\label{(22)}
\end{equation}
As $\cos\delta$ increases, $x_s/x_d$ monotonously
increases.  The present constraints on $\cos\delta$ and
$|V_{13}/V_{23}|$ give $10\lsim x_s/x_d\lsim 56$.
It is worth emphasizing that the three $CP$ asymmetries
and the ratio $x_s/x_d$ only depend on
$\cos\delta$ and $|V_{13}/V_{23}|$.

\begin{table}
\caption{The ranges of $\sin 2\phi_1$,
          $\sin 2\phi_2$, $\sin 2\phi_3$, and
          $x_s/x_d$ predicted by the SSM and the SM:
          (a) $|V_{13}/V_{23}|=0.08$, $|V_{23}|=0.04$,
          (b) $|V_{13}/V_{23}|=0.08$, $|V_{23}|=0.044$,
          (c) $|V_{13}/V_{23}|=0.1$, $|V_{23}|=0.04$.}
\begin{center}
\begin{tabular}{c|c|c|c|c}
\hline
 & $\sin 2\phi_1$ & $\sin 2\phi_2$ & $\sin 2\phi_3$ & $x_s/x_d$ \\
\hline
 (a) SSM & (0.55, 0.66) & $(-0.96, 0.74)$ & $(-0.18, 1.00)$
					    & (17, 36) \\
  SM & (0.61, 0.66) & (0.15, 0.74) & $(-0.18, 0.54)$
					     & (17, 21)  \\
\hline
 (b) SSM & (0.46, 0.66) &  $(-1.00, 0.31)$ & (0.39, 1.00)
					     & (20, 40) \\
  SM & (0.63, 0.66) & $(-0.65, 0.31)$  & (0.39, 0.98)
					     & (20, 30) \\
\hline
 (c) SSM & (0.56, 0.80) & $(-1.00, 0.73)$  & (0.02, 1.00)
					     & (17, 53) \\
  SM & (0.74, 0.80) & $(-0.15, 0.73)$  & (0.02, 0.88)
					     & (17, 27) \\
\hline
\end{tabular}
\end{center}
\label{cpasy}
\end{table}
     In Table \ref{cpasy} we give the predicted values of
$\sin 2\phi_i$
and $x_s/x_d$ in the SSM and in the SM for
$(|V_{13}/V_{23}|, |V_{23}|)=(0.08, 0.04)$ (a),
(0.08, 0.044) (b), (0.1, 0.04) (c).
In each case there are wide ranges of $\sin 2\phi_2$,
$\sin 2\phi_3$, and/or $x_s/x_d$ which are possible
only in the SSM.
If the experimental values are found in these ranges, then
this would indicate that $R>1$, with significant new
contributions to \BB\ and \KK\ mixings arising from
the SSM.
For case (a)
the possible experimental results
$\sin 2\phi_2 <0$, $\sin 2\phi_3\sim 1$,
and $x_s/x_d\sim 30$
suggest the SSM effects.
Note, however, that the predicted ranges vary with
the values of $|V_{13}/V_{23}|$ and $|V_{23}|$.  More
precise measurements for these quantities are necessary
to make predictions definitely.

\section{Summary}

     In the SSM there exist several new interactions
which can induce FCNC processes.  We have discussed
their effects on \BdBd\ and \KK\ mixings.
These mixings receive contributions from box diagrams in which
charginos and up-type squarks, or charged Higgs bosons
and up-type quarks are exchanged.  We have calculated the
ratio $R$ of the SSM contribution to the contribution
in the SM.  The new SSM contributions interfere constructively
with the standard $W$-boson contribution.  The ratio $R$
is sizably larger than unity,
if $\tb$ has a value around unity
and a chargino, a $t$-squark, and/or a charged Higgs boson
are not much heavier than 100 GeV.

     The enhanced SSM contributions to \BdBd\ and \KK\
mixings make the $CP$-violating
phase $\delta$ of the CKM matrix have a value different
from the SM prediction.
We have
discussed the ranges of $\cos\delta$ and $R$
which are derived from $x_d$ and $\e$.  The present
uncertainties in $|V_{13}/V_{23}|$, $|V_{23}|$,
$f_{B_d}\sqrt{B_{B_d}}$, and $B_K$ are still large,
and the allowed ranges for $\cos\delta$ and $R$
vary with the values of those quantities.
However,
if the present experimental central values for
$|V_{13}/V_{23}|$ and $|V_{23}|$ are in the vicinities
of actual values, $\cos\delta$ and $R$ should lie in the
ranges $(-0.1, 0.8)$ and (1.0, 2.1), respectively, while
a constraint $-0.1\lsim\cos\delta\lsim 0.3$ is obtained for the
SM, which corresponds to $R=1$.
This $CP$-violating phase $\delta$ can be
probed by $CP$ asymmetries in $B^0$-meson decays
and amount of \BsBs\ mixing.
We have shown the possibility that the measurements
of $\sin 2\phi_1$, $\sin 2\phi_2$,
$\sin 2\phi_3$, and $x_s/x_d$ disclose values
of $\cos\delta$ and $R$ which are not allowed in the SM,
thereby implicating the existence of supersymmetry.

     The SSM interactions mediated by
the charginos and the charged Higgs bosons
could also sizably contribute to radiative
$B$-meson decay \cite{bsg}.  The chargino contribution interferes
with the standard $W$-boson contribution either constructively
or destructively depending on the parameter values,
while the charged Higgs boson contribution interferes
constructively.
Compared with the SM prediction, the branching ratio
of $B\rightarrow X_s\gamma$ becomes either enhanced or
reduced, if $\tb$ is much larger than unity or the relevant
particles are not heavy.
Radiative $B$-meson decay therefore provides
information on the SSM complimentarily to \BB\ and \KK\
mixings.

\section*{Acknowledgement}

     This article is based on the works in collaboration
with G.C. Branco, G.C. Cho, and the late Y. Kizukuri.


\begin{thebibliography}{99}

\bibitem{ellis} J. Ellis and D.V. Nanopoulos, Phys. Lett.
            {\bf 110B} (1982) 44;        \\
          R. Barbieri and R. Gatto,  Phys. Lett.
            {\bf 110B} (1982) 211; \\
          T. Inami and C.S. Lim,  Nucl. Phys. {\bf B207}
          (1982) 533;   \\
           M.J. Duncan,  Nucl. Phys. {\bf B221} (1983) 285; \\
         J.F. Donoghue, H.P. Nilles, and D. Wyler,
            Phys. Lett. {\bf 128B} (1983) 55.
\bibitem{wise} L.F. Abbott, P. Sikivie, and M.B. Wise,
            Phys. Rev. {\bf D21} (1980) 1393.
\bibitem{BB} G.C. Branco, G.C. Cho, Y. Kizukuri, and N. Oshimo,
             Phys. Lett. {\bf B337} (1994) 316.
\bibitem{bb2} G.C. Branco, G.C. Cho, Y. Kizukuri, and N. Oshimo,
	       Nucl. Phys. {\bf B449} (1995) 483.
\bibitem{SUSY} For reviews, see e.g.
           H.P. Nilles,  Phys. Rep. {\bf 110} (1984) 1;  \\
    H.E. Haber and G.L. Kane,  Phys. Rep. {\bf 117} (1985) 75.
\bibitem{bsg}N. Oshimo,  Nucl. Phys. {\bf B404} (1993) 20.
\bibitem{SUSYBB}T. Kurimoto,  Phys. Rev. {\bf D39} (1989)
             3447;  \\
           S. Bertolini, F. Borzumati, A. Masiero, and G. Ridolfi,
            Nucl. Phys. {\bf B353} (1991) 591.
\bibitem{EDM}Y. Kizukuri and N. Oshimo, Phys. Rev.
	   {\bf D45} (1992) 1806; Phys. Rev. {\bf D46} (1992) 3025.
\bibitem{FCNC}For reviews, see e.g.
           J.F. Donoghue, B.R. Holstein, and G. Valencia,
            Int. J. Mod. Phys. {\bf A2} (1987) 319; \\
           W. Grimus,  Fortschr. Phys. {\bf 36} (1988) 201; \\
           P.J. Franzini,  Phys. Rep. {\bf 173} (1989) 1.
\bibitem{inami} T. Inami and C. S. Lim,
                Prog. Theor. Phys. {\bf 65} (1981) 297 (E: 1772).
\bibitem{top} CDF Collaboration,  Phys. Rev. {\bf D50} (1994)
           2966; Phys. Rev. Lett. {\bf 73} (1994) 225.
\bibitem{PDG}Particle Data Group,  Phys. Rev. {\bf D50}
          (1994) 1173.
\bibitem{CKM}J.R. Patterson, in Proc. of the XXVII
           International Conference on High Energy Physics,
           eds. P.J. Bussey and I.G. Knowles (IOP Publishing,
	   Bristol and Philadelphia, 1995).
\bibitem{Klattice}P.B. Mackenzie, in Proc. of the XVI International
           Symposium on Lepton and Photon Interactions,
           eds. P. Drell and D. Rubin (AIP, New York, 1994).
\bibitem{KN} W.A. Bardeen, A.J. Buras, and J.-M. Gerard, Phys. Lett.
             {\bf B211} (1988) 343.
\bibitem{Blattice}A. Abada et al., Nucl. Phys. {\bf B376}
          (1992) 172; \\
	   A. Abada, in Proc. of the XXVII
           International Conference on High Energy Physics,
           eds. P.J. Bussey and I.G. Knowles (IOP Publishing,
	   Bristol and Philadelphia, 1995).
\bibitem{QCD}W.A. Kaufman, H. Steger, and Y.-P. Yao, Mod. Phys. Lett.
             {\bf A3} (1988) 1479; \\
 A. Datta, J. Fr\"ohlich, and E.A. Paschos, Z. Phys. {\bf C46}
             (1990) 63; \\
             J.M. Flynn, Mod. Phys. Lett. {\bf A5} (1990) 877;  \\
 A.J. Buras, M. Jamin, and P.H. Weisz, Nucl. Phys. {\bf B347}
             (1990) 491; \\
 S. Herrlich and U. Nierste, Nucl. Phys. {\bf B419} (1994) 292.
\bibitem{stop}G.C. Cho, Y. Kizukuri, and N. Oshimo,
              TKU-HEP 95/02, OCHA-PP-62.
\bibitem{CPasy} For reviews, see e.g.
           I.I. Bigi, V.A. Khoze, N.G. Uraltsev, and A.I.
           Sanda, in {\it CP violation}, ed. C. Jarlskog
           (World Scientific, Singapore, 1989);  \\
 Y. Nir and H.R. Quinn, Ann. Rev. Nucl. Part. Sci. {\bf 42}
           (1992) 211.
\bibitem{gilman}P. Krawczyk, D. London, R.D. Peccei, and H. Steger,
                Nucl. Phys. {\bf B307} (1988) 19;  \\
              C.O. Dib, I. Dunietz, F.J. Gilman, and Y. Nir,
                Phys. Rev. {\bf D41} (1990) 1522; \\
                M. Lusignoli, L. Maiani, G. Martinelli, and
                L. Reina, Nucl. Phys. {\bf B369} (1992) 139; \\
                A.J. Buras, M.E. Lautenbacher, and G. Ostermaier,
                Phys. Rev. {\bf D50} (1994) 3433; \\
       A. Ali and D. London, Z. Phys. {\bf C65} (1995) 431.
\end{thebibliography}
\end{document}